\documentclass[12pt]{iopart}
\usepackage{amssymb}
\usepackage{graphicx}

\begin{document}

\title{L\'evy ratchets with dichotomic random flashing}

\author{S A Ib\'a\~nez, A B Kolton, S Risau-Gusman and S Bouzat}
\address{Consejo Nacional de Investigaciones Cient\'{\i}ficas y T\'ecnicas, \\
Centro At\'omico Bariloche (CNEA), (8400) Bariloche, R\'{\i}o Negro, Argentina.}
\ead{bouzat@cab.cnea.gov.ar}


\begin{abstract}
Additive symmetric L\'evy noise can induce directed transport of overdamped particles in
a static asymmetric potential. We study, numerically and analytically, the effect of an
additional dichotomous random flashing in such L\'evy ratchet system. For this purpose we
analyze and solve the corresponding fractional Fokker-Planck equations and we check the
results with Langevin simulations. We study the behavior of the current as function of
the stability index of the L\'evy noise, the noise intensity and the flashing parameters.
We find that flashing allows both to enhance and diminish in a broad range the static
L\'evy ratchet current, depending on the frequencies and asymmetry of the multiplicative
dichotomous noise, and on the additive L\'evy noise parameters. Our results thus extend
those for dichotomous flashing ratchets with Gaussian noise to the case of broadly
distributed noises.
\end{abstract}

\maketitle

\section{Introduction}\label{sec_intro}

The study of noise induced transport in anisotropic spatially
periodic systems is a relevant subject for many problems in physics
\cite{general1,rmp09} and biology \cite{biology}, and is also
acquiring an increasing interest due to possible applications in the
development of technological devices \cite{rmp09}.

In general, the basic models consider a particle in an external
spatially periodic potential with broken symmetry (the {\em
ratchet-potential}), and subject to an additional signal or source
of fluctuations which generates the nonequilibrium condition
necessary for the emergence of directional transport \cite{rmp09}.
Thermal noise is also usually considered, although depending on the
nature of the nonequilibrium forces it may be not a fundamental
ingredient. Ratchet models can be classified in two main classes
according to the way in which the non equilibrium forcing affects
the particle dynamics. Simple additive nonequilibrium forces lead to
the so called {\em rocking ratchets}, while, when the signal enters
as a multiplicative modulation of the ratchet potential, we speak of
{\em flashing ratchets}. Systems combining the two kind of forcings
can also
been considered~\cite{marchesoni}. 

Recently, two simultaneous papers~\cite{dybiec,negrete} have shown that a minimal setup
for producing directional transport is obtained by considering a simple static ratchet
potential and an additive white symmetric L\'evy \cite{dybiec,negrete,LevyMath} noise as
the only two ingredients. Here we will refer to such system as the {\em static} or {\em
non flashing} L\'evy ratchet. The preferred direction of motion for such system is found
to be towards the steepest slope of the potential. The effect was explained in
\cite{dybiec} as a consequence of the large L\'evy jumps, which lead the particles to the
flatter zones of the potential with larger probability than to the steeper zones. In the
limit case in which the stability parameter $\alpha$ \cite{dybiec,negrete} defining the
L\'evy noise is set to be equal to $2$, the L\'evy distribution becomes
Gaussian\cite{dybiec,negrete} and the equilibrium situation with vanishing current is
recovered.

After the mentioned pioneering papers on the static L\'evy ratchet, several works have
appeared providing further analysis of the system and studying different generalizations.
In \cite{dybiec2} it has been shown that an inversion of current can be obtained by
considering a time periodic modulation of the chirality of the L\'evy noise. Reference
\cite{bouzatPhysA} studies inertial effects and propose a way of measuring the
rectification efficiency in L\'evy ratchets. In \cite{weaknoise}, the weak noise limit is
analyzed. In \cite{negrete2010}, the related problem of the spatially tempered fractional
Fokker-Planck equation is studied. Reference \cite{aclevy} analyzes the competition
between L\'evy forcing and a periodic a.c. driving  in a ratchet system, while
\cite{sbudlevy} studies the coexistence of L\'evy flights and subdiffusion. The
increasing interest in ratchet systems influenced by L\'evy noises is due to various
facts. Firstly, there is an intrinsic theoretical significance in the generalizations of
previous ratchet models to account for cases where fluctuations present long tailed
probability distribution, giving rise to anomalously large particle displacements.
Moreover, since L\'evy noises may induce divergencies in the moments of the velocity
distributions \cite{dybiec}, there is an additional challenge in providing the
appropriate quantities to measure currents \cite{dybiec,negrete2010}, particle dispersion
\cite{dybiec,negrete2010} and efficiencies \cite{bouzatPhysA}. From another point of
view, L\'evy ratchets may be of direct interest for problems in magnetically confined
fusion plasmas \cite{negrete,bouzatPhysA,impurity}. L\'evy ratchets might be also of
interest for atomic transport in dissipative optical lattices. Cold atom ratchets have
been indeed recently studied
experimentally~\cite{phil,gommers05a,gommers05b,quasip,gating} and
theoretically~\cite{brown,kolton}, whereas anomalous dynamics described by L\'evy
statistics were also reported for the same system~\cite{katori,lutz}.\\

In this paper we study the dynamics of a ratchet system influenced
by two different signals inducing non equilibrium conditions: a
dichotomic multiplicative noise and an additive L\'evy forcing. We
analyze both the Langevin and Fokker-Plank approaches, focusing
mainly on the latter. In different limit situations, the model
generalizes various previous systems found in the literature. As we
will see, for fast, slow and null flashing we recover different
versions of the static L\'evy ratchet, while for $\alpha\to 2$ we
recover a version of the well known flashing ratchet with thermal
noise \cite{general1,rmp09}. In particular we find that flashing
allows both to enhance and diminish in a broad range the static
ratchet current depending on the frequencies of flashing. Our
results thus extend those for dichotomous flashing ratchets with
gaussian noise to the case of broadly distributed noise.

It is worth mentioning that a dichotomous flashing ratchet with
L\'evy noise was shortly analyzed in the past as part of other
studies on the influence of supperdiffusion on directional motion
\cite{Bao}. However, only heuristic arguments were given by
considering an adiabatic approximation which assumes relatively slow
flashing. Interestingly, as the study is previous to the findings in
\cite{dybiec} and \cite{negrete}, the contribution to directional
transport during the "on" stage of the flashing was naturally
ignored.

The organization of the paper is as follows. In section \ref{sec_model} we
introduce the model. In section \ref{sec_sol} we give the solution of the
Fokker-Planck equation. In section \ref{sec_limitsituations}  we analyze some limit
situations which connect our results with previous results for other
systems studied in the literature. In section \ref{sec_results} we present a
detailed numerical analysis of the results considering a particular
but standard ratchet potential. Section \ref{sec_conclusion} is devoted to our
conclusions and some final remarks.

\section{Model}\label{sec_model}
\begin{figure}[ht!]
\centering
    \includegraphics[width=0.45\textwidth]{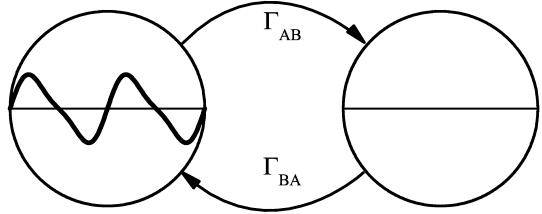}
    \caption{Scheme of the L\'evy ratchet model with dichotomic random
flashing.}
    \label{Fig_0a}
\end{figure}

We consider the one-dimensional overdamped motion of a
particle at position $X(t)$ subject to a randomly
fluctuating ratchet potential $f(t)V(x)$,
and to an additive L\'evy random force $\xi(t)$. This situation is
described by the Langevin equation
\begin{equation}\label{Ec_Langevin}
\frac{d X}{d t}=-f(t) V'(X)+\xi(t).
\end{equation}
We consider $\xi(t)$ as a symmetric $\alpha$-stable L\'evy noise with $1 < \alpha \leq 2$
and intensity $\chi$, described by the characteristic function $\langle \exp[ik
\int_t^{t+\Delta t} \xi(t') dt'] \rangle = \exp[-\chi|k|^\alpha \Delta
t]$~\cite{dybiec,negrete,LevyMath}. We consider a ratchet potential with spatial period
$L$ and amplitude $\Delta V$, and the multiplicative noise $f(t)$ as a dichotomous Markov
process \cite{aaa,bbb} taking the value of $1$ (mode A) or $0$ (mode B), switching between these modes
at rates $\Gamma_{AB}$ and $\Gamma_{BA}$, respectively. Equation (\ref{Ec_Langevin}),
thus, models a particle in a ratchet potential $V(x)$ that switches between `on'  and
`off' states with transition rates $\Gamma_{AB}$ and $\Gamma_{BA}$ respectively (see
Figure \ref{Fig_0a}).
We are interested in the steady-state directional transport that can
be generated in this system. This can be quantified by the long-time
limit of the average velocity $\langle dX/dt \rangle$, where
$\langle...\rangle$ stands for the average over all possible
realizations of $\xi(t)$ and $f(t)$. Note that we only consider the
domain $1<\alpha\leq 2$ since it was shown that transport is highly
inefficient in L\'evy ratchets for $\alpha < 1$ due to the
divergencies of the mean value of L\'evy distributions
\cite{dybiec,bouzatPhysA}. Moreover, the analysis in such parameter
region demands special care for defining the mean velocity and
current \cite{dybiec}. Note that, our parameter region of interest
includes the case $\alpha=2$, for which the L\'evy noise becomes
Gaussian white noise and (\ref{Ec_Langevin}) describes the
dichotomous flashing ratchet studied in \cite{astumian_bier}. On
the other hand, for $\Gamma_{AB}=0$, equation (\ref{Ec_Langevin})
corresponds to the static or non flashing L\'evy ratchet mentioned
in the introduction and recently analyzed in \cite{dybiec,negrete}.

The Langevin equation (\ref{Ec_Langevin}) was solved by using a
standard algorithm (see \cite{dybiec} and references in
\cite{dybiec} and \cite{negrete}) for which the numbers distributed
according to L\'evy laws were obtained by the method explained in
\cite{numeric}.

As mentioned in the introduction, most of our analysis will be done using the
Fokker-Planck approach to the problem. In order to present it, we first consider a more
general situation in which the potential in state B can be non vanishing, and we
introduce the probability densities $P^A=P^A(x,t)$ and $P^B=P^B(x,t)$ of finding the
particle at position $x$ and time $t$ in the A and B modes respectively. For such a
system we have the following coupled Fokker-Planck equations for the probability
densities
\begin{equation} \label{Ec_FokkerPlanck}
    \begin{array}{lcl}
        \partial_t P^A &=& \partial_x (P^A \partial_x V^A) + \chi
\partial_x^{\alpha} P^A + \Gamma_{BA} P^B - \Gamma_{AB} P^A\\
        \partial_t P^B &=& \partial_x (P^B \partial_x V^B) + \chi
\partial_x^{\alpha} P^B + \Gamma_{AB} P^A - \Gamma_{BA} P^B. \\
    \end{array}
\end{equation}
Here, $\partial_x^{\alpha} \equiv \partial^{\alpha}/\partial
|x|^{\alpha}$ stands for the Riesz-Feller
fractional derivative of order $\alpha$~\cite{podlubny},
while $V^A(x)$ and $V^B(x)$ are the external potentials
considered in states A and B. For $V^B(x)=0$, the system
(\ref{Ec_FokkerPlanck}) is equivalent to the Langevin equation
(\ref{Ec_Langevin}). We will limit to such a case when performing
our analysis of transport and presenting our numerical results.
However, the method of solution of the Fokker-Planck equation will
be introduced (up to some point) for an arbitrary potential
$V^B(x)$, since this leads to more general and symmetrical
expressions.

Within the Fokker-Planck framework, the probability current $J(x,t)
= \langle (dX(t)/dt) \delta(x-X(t)) \rangle$ is obtained from the
continuity equation $\partial_t P = -\partial_x J$, where $P \equiv
P^A + P^B \equiv \langle \delta(x-X(t)) \rangle$ is the probability
density regardless of the mode. In the long-time limit $\partial_t P
=\partial_t P^A = \partial_t P^B = 0$ and the current reaches the
spatially constant value $J$ characterizing the average directed
motion or ratchet response. Since the steady-state solutions we are
interested in, $P^A(x)$ and $P^B(x)$, are periodic with the period
$L$, we can direclty solve (\ref{Ec_FokkerPlanck}) for
$\partial_t P^A = \partial_t P^B = 0$ in the interval $0 \leq x \leq L$ with periodic
boundary conditions. The steady-state current is then simply related
to the average velocity in the Langevin description, $J=\langle
dX(t)/dt\rangle/L$.

\section{Solution of the Fokker-Planck Equation}\label{sec_sol}

We exploit the fact that the steady-state solution we seek is
periodic with the period $L$ of the ratchet potential by using
discrete Fourier transforms, $f(x)=\sum_q \tilde{f}_q \exp(\rmi q x)$,
where $q \equiv 2\pi n_q/L$ with $n_q$ an integer and
$\tilde{f}_k=\int_0^L (dx/L)\; f(x) \exp(-\rmi k x)$ denotes the
correponding Fourier amplitude. In particular, the prescription
$\widetilde{[\partial_x^{\alpha} f(x)]_k}=-|k|^{\alpha} \tilde{f}_k$
for the Fourier transform of the fractional derivative, valid in an
infinite or periodic support, greatly simplifies our method as it
avoids the complications due to the nonlocal nature of the
fractional Laplacian operator that can arise for different boundary
conditions that break translational invariance in bounded
domains \cite{rosso}. Fourier transforming
(\ref{Ec_FokkerPlanck}) gives
\begin{equation}
    \begin{array}{lcl}
        \partial_t {\tilde P}^A_k &=&  -k \sum_q  q
\tilde{V}^A_q \tilde{P}^A_{k-q} - \chi |k|^{\alpha} \tilde{P}^A_k +
                    \Gamma_{BA} \tilde{P}^B_k - \Gamma_{AB}
\tilde{P}^A_k \\
        \partial_t {\tilde P}^B_k &=&  -k \sum_q  q
\tilde{V}^B_q \tilde{P}^B_{k-q} - \chi |k|^{\alpha} \tilde{P}^B_k +
                    \Gamma_{AB} \tilde{P}^A_k - \Gamma_{BA}
\tilde{P}^B_k,
    \end{array}
\label{Ec_FTFPeqs}
\end{equation}
where we have used (and use from now on) indistintevily the wave
vector name, eg $k$, to denote both its value, $k=2\pi n_k/L$ and
its associated integer, $n_k$, to avoid excesive notation. By
construction equation (\ref{Ec_FTFPeqs}) only admit periodic initial
conditions with period $L$. This is not important as we will be
interested in the steady-state limit $\partial_t {\tilde
P}^{A,B}_k=0$, whose solution is unique and thus independent of the
initial condition. Equation (\ref{Ec_FTFPeqs}) must be solved with the
constraint $\tilde{P}_0^A+\tilde{P}_0^B=1$, arising from the
normalization of the entire probability function $\int_0^L dx\; P =
1$. This leads inmediately to the zero-mode steady-state solution
\begin{equation}
    \begin{array}{lcl}
\tilde{P}_0^A=\Gamma_{BA}/(\Gamma_{BA}+\Gamma_{AB})\\
\tilde{P}_0^B=\Gamma_{AB}/(\Gamma_{BA}+\Gamma_{AB}),
    \end{array}
\label{Ec_FPzeromode}
\end{equation}
which physically simply states that the probability of finding the
system in the A and B modes regardless of the particle position, are
controlled by the average fraction of time spent on each mode,
$\tau_A/(\tau_A+\tau_B)$ and $\tau_B/(\tau_A+\tau_B)$ respectively,
where $\tau_A/\tau_B = \Gamma_{BA}/\Gamma_{AB}$.

Using the zero-mode solution of (\ref{Ec_FPzeromode}) the system
(\ref{Ec_FTFPeqs}) can be written as an inhomogeneous system
for all the components $\tilde{P}_k^A$ and $\tilde{P}_k^B$ with $k
\neq 0$,
\begin{equation}
    \begin{array}{lcl}
        -k \sum_{q \neq k}  q
\tilde{V}^A_q \tilde{P}^A_{k-q} - \chi |k|^{\alpha} \tilde{P}^A_k +
                    \Gamma_{BA} \tilde{P}^B_k - \Gamma_{AB}
\tilde{P}^A_k &=& k^2 \tilde{V}^A_k \delta \\
        -k \sum_{q \neq k}  q
\tilde{V}^B_q \tilde{P}^B_{k-q} - \chi |k|^{\alpha} \tilde{P}^B_k +
                    \Gamma_{AB} \tilde{P}^A_k - \Gamma_{BA}
\tilde{P}^B_k &=& k^2 \tilde{V}^B_k (1-\delta),
    \end{array}
\label{Ec_FTFPeqs2}
\end{equation}
where we have defined the duty ratio or fraction of time spent in
the A mode, $\delta \equiv \Gamma_{BA}/(\Gamma_{BA}+\Gamma_{AB})$.
In the particular case of a potential that is stochastically
switched on and off, we have $V^A(x) = V(x)$ and $V^B(x) \equiv 0$
and thus $\tilde{V}^B=0$. Using the second equation in the system of
Eqs. (\ref{Ec_FTFPeqs2}) we obtain,
\begin{equation}
 \tilde{P}_k^B = \frac{\Gamma_{AB}}{\chi |k|^{\alpha} + \Gamma_{BA}}  \tilde{P}_k^A,
\end{equation}
and therefore the problem is reduced to solve,
\begin{equation}
\label{Ec_unasola} -k \sum_{q \neq k}  q \tilde{V}_q
\tilde{P}^A_{k-q} - \chi |k|^{\alpha}
\left(1+\frac{\Gamma_{AB}}{\chi |k|^{\alpha} + \Gamma_{BA}}\right)
\tilde{P}^A_k = k^2 \tilde{V}_k \delta\;\;\;\;\;\;\; (k\neq 0)
\end{equation}
with $\tilde{P}_0^A=\delta = 1-\tilde{P}_0^B$, from (\ref{Ec_FPzeromode}).
In a more compact matrix notation we have
\begin{equation}
(\mathbf{M} + \mathbf{D}) \tilde{P}^A = C \delta, \\
\label{Ec_FTFPeqsmat}
\end{equation}
where the components of the
matrices $\mathbf{M}$, $\mathbf{D}$ and the vector $\mathbf{C}$
are defined as, for $k\neq0$,
\begin{equation}\label{Ec_Matrix}
    \begin{array}{lcl}
  M_{kq}= -k(k-q) \tilde{V}_{k-q},\\
  D_{kq}= - \chi |k|^{\alpha} \left(1+\frac{\Gamma_{AB}}{\chi
|k|^{\alpha} + \Gamma_{BA}}\right)\delta_{kq}, \\
  C_k \equiv -M_{k0} = k^2 \tilde{V}_{k}.
    \end{array}
\end{equation}
In this way, $\mathbf{M}$ and $C$ depend only on the Fourier
components of the ratchet potencial $V(x)$, while the diagonal
matrix $\mathbf{M}$ concentrates, for fixed $\delta$,
all the dependence with the
external parameters $\Gamma_{AB}$, $\Gamma_{BA}$, $\chi$ and
$\alpha$.

Finally, in order to compute the current we consider the
Fourier-transformed continuity equation
\begin{equation}
\label{contfou} \partial_t \tilde{P_k}(t) = -\rmi k \tilde{J}_k(t),
\end{equation}
where $\tilde{P_k}(t)=\tilde{P}^A_{k}(t)+\tilde{P}^B_{k}(t)$. Adding
the two equations (\ref{Ec_FTFPeqs}) together, we can identify:
\begin{equation}
\label{jotaka}
 \tilde{J}_k = - \rmi \sum_q  q \tilde{V}_q \tilde{P}^A_{k-q} - \rmi \chi |k|^{\alpha-1} \mbox{sgn}(k) (\tilde{P}^A_k+\tilde{P}^B_k).
\end{equation}
As indicated in the previous section, the current in the stationary
regime is a constant. This can be seen from (\ref{contfou}),
which for $\partial_t \tilde{P_k}(t)=0$ implies $J_k=0$ for $k\neq
0$. Thus, assuming $\alpha>1$, the stationary current is simply
\begin{equation}
J \equiv J_0= - \rmi \sum_q  q \tilde{V}_q \tilde{P}^A_{-q}.
\label{Ec_current}
\end{equation}

Note that the solution for the general case $V^B(x)\neq 0$ demands
dealing with the whole system (\ref{Ec_FTFPeqs2}).
Clearly, although the procedure is a little bit more intricate, the
equations can also be written in matrix form and solved for the
variables $\tilde{P}_k^A$ and $\tilde{P}_k^B$, enabling the
calculation of the current.

\section{Dimensional analysis, relevant parameters and limit situations}\label{sec_limitsituations}

Considering a potential $V\left(x\right)$ of amplitude $\Delta V$ and period $L$,
a straight forward dimensional analysis of (\ref{Ec_unasola})
shows that the Fokker-Planck solutions for the probability
distribution and current are of the form
\begin{eqnarray}
\label{scaling} P &=& f\left(\frac{x}{L};\delta,\alpha,\frac{\Gamma_{AB} L^2}{\Delta V},
\frac{\chi L^{2-\alpha}}{\Delta V} \right), \nonumber \\
J &=& g\left(\alpha,\delta,\frac{\Gamma_{AB} L^2}{\Delta
V},\frac{\chi L^{2-\alpha}}{\Delta V} \right) \frac{\Delta V}{L},
\end{eqnarray}
where $f$ and $g$ are dimensionless functions. Thus, there are essentially four relevant
parameters.
Note that the scaled `on' rate ${\Gamma_{BA} L^2}/{\Delta V}$ can be
considered as a relevant parameter instead of $\delta$ or instead of
${\Gamma_{AB}L^2}/{\Delta V}$. In the following we will speak alternatively of the three parameters depending on the feature we want to stress. Note that for $\delta=1$, (\ref{Ec_unasola}) reduces to the equation for the standard, non flashing, L\'evy ratchet
\cite{negrete,negrete2010}.

The relaxation time for a particle in the potential $V(x)$ is
expected to be a relevant characteristic time of the system.
According to the deterministic part of (\ref{Ec_Langevin}),
considering a typical length $L$ and a velocity $(\Delta V)/L$, we
get $L^2/\Delta V$ as the typical relaxation time. It is interesting
to note thus, that the scaled parameters ${\Gamma_{AB} L^2}/{\Delta
V}$ and ${\Gamma_{BA} L^2}/{\Delta V}$ are simply measures of the
switching rates in the time scale of the inverse of the relaxation
time. Equivalently, $L^{\alpha}/\chi$ can be interpreted as a
super-diffusion time which in the solution of the Fokker-Planck
equation appears weighted to the relaxation time. Note that in the
Gaussian limit we get the usual expression for the diffusion time
$\equiv L^{2}/\chi$.

By modifying $\Gamma_{AB}$ and $\Gamma_{BA}$ but keeping
constant their ratio (i.e. keeping $\delta$ constant), we can analytically study the
limits of infinitely slow and infinitely fast switching without changing the relative
residence times. Using (\ref{Ec_unasola}) the calculation of these limits is very simple
and instructive.

\subsection{Infinitely slow switching}
In the infinitely slow switching case we have $\Gamma_{AB} \to 0$ and
$\Gamma_{BA} \to 0$, keeping $\delta$ constant.
Replacing this in (\ref{Ec_unasola}) and using (\ref{Ec_FPzeromode}), we can write
\begin{equation}
    \begin{array}{lcl}
-k \sum_{q \neq k}  q \tilde{V}_q
(\tilde{P}^A_{k-q}/\delta) - \chi |k|^{\alpha}
(\tilde{P}^A_k/\delta) = k^2 \tilde{V}_k\;\;\;\;\;\;\;\;\; (k\neq 0) \\
(\tilde{P}^A_{0}/\delta) = 1.
    \end{array}
\end{equation}
These equations are identical to the non-flashing L\'evy ratchet
equations for $\tilde{P}_k^A/\delta$. If we call $P_{st}(V)$ the
real-space static L\'evy ratchet solution in a potential $V$, we
have, in this slow switching limit,
\begin{equation}
    \begin{array}{lcl}
P^A_{slow} = P_{st}(V) \, \delta, \\
J_{slow} = J_{st}(V) \, \delta,
    \end{array}
\label{Ec_slowsolution}
\end{equation}
where the second equation follows from (\ref{Ec_current}). Intuitively, the idea is that
the rates are so large that the system reaches the steady state when the potential is
`on' and when it is `off', and having the transients between these modes a negligible
contribution, the total current is just an average of the two corresponding steady state
currents. According to (\ref{Ec_unasola}) the result is expected to be a very good
approximation when the following relations hold between the scaled parameters:
${\Gamma_{AB} L^2}/{\Delta V}\ll {\chi L^{2-\alpha}}/{\Delta V}$ and
${\Gamma_{BA} L^2}/{\Delta V} \ll {\chi L^{2-\alpha}}/{\Delta V}$ (i.e.
$\Gamma_{AB}\ll \chi/L^\alpha$ and $\Gamma_{BA}\ll \chi/L^\alpha$). We will check this in
section \ref{sec_results} when analyzing transport for the case of a standard ratchet potential.

\subsection{Infinitely fast switching}
In the infinitely fast switching case we have $\Gamma_{AB} \to
\infty$ and $\Gamma_{BA} \to \infty$, keeping $\delta$ constant.
Replacing this in (\ref{Ec_unasola}) the bracketed term becomes
$1/\delta$. We can thus write,
\begin{equation}
    \begin{array}{lcl}
 -k \sum_{q \neq k}  q \tilde{V}_q \delta
(\tilde{P}^A_{k-q}/\delta) - \chi |k|^{\alpha}
(\tilde{P}^A_k/\delta) = k^2 \tilde{V}_k \delta\;\;\;\;\;\;\; (k\neq 0) \\
(\tilde{P}^A_{0}/\delta) = 1.
    \end{array}
\end{equation}
which again correspond to the non-flashing ratchet equations for the
variable $\tilde{P}^A_{k-q}/\delta$ but in the renormalized
potential $V \, \delta$. Thus, the solution in this fast-switching
limit is
\begin{equation}
    \begin{array}{lcl}
P^A_{fast} = P_{st}(V \, \delta ) \, \delta \\
J_{fast} = J_{st}(V \, \delta)
    \end{array}
\label{Ec_fastsolution}
\end{equation}
In this case, the current for the flashing L\'evy ratchet is the
same as for a system where the same potential is always `on'but with
an intensity rescaled by $\delta$. This is intuitively evident: the
potential is being switched so fastly that the particle can only
`feel' the temporal average of the potential (i.e. $V(x) \, \delta
$).
According to (\ref{Ec_unasola}) the limit results in (\ref{Ec_fastsolution})
are expected to be valid for $\Gamma_{AB}\gg \chi/L^\alpha$ and
$\Gamma_{BA}\gg \chi/L^\alpha$. We have verified
this using a standard differentiable ratchet potential, as we will
show in section \ref{sec_results}.

Note however that, at variance with what happens in the slow limit,
the fast limit is not valid for all the Fourier modes.
The denominator of the bracketed term in (\ref{Ec_unasola}) shows that
the fast approximation is only valid for Fourier modes that satisfy
$|k|^\alpha \ll \Gamma_{AB} / \chi$. Nevertheless, if the potential
is an analytic function the probability density function, $P(x)$
should also be an analytic function, which implies that the
coefficients of its Fourier expansion decay exponentially with $k$.
Therefore we expect that the higher Fourier modes that are not well
approximated in the fast limit do not change the overall limit of
the probability density.

\subsection{Perturbative analysis for fast and slow switching}
Using a simple perturbation analysis in (\ref{Ec_unasola}) we can
obtain the lowest order corrections to the infinitely slow and fast
switching limits described above, with $\Gamma_{AB}$ and
$\Gamma_{AB}^{-1}$ the small parameters, respectively, for a fixed value
of $\delta$.

Approaching the infinitely slow limit a natural ansatz is to propose, $P^A (x) \sim
P^A_{slow} (x) + \Gamma_{AB} W(x)$, with $P^A_{slow}$
given by (\ref{Ec_slowsolution}).
Inserting this ansatz in (\ref{Ec_unasola}) we get the following result for the
fourier-transformed first order correction for $k\neq 0$,
\begin{equation}
-k\sum_{q \neq k} q {\tilde V}_q {\tilde W}_{k-q}-\chi
|k|^{\alpha}{\tilde W}_k = [{\tilde P^A_{slow}}]_k.
\label{Ec_1stslow}
\end{equation}
For various particular potentials we confirm numerically that (\ref{Ec_1stslow}) has
solutions with finite currents. As we will see, this is also consistent with the slow
switching behaviour found in the full numerical solution discussed in the next section.
We therefore conclude that in the absence of particular symmetries, the first order
correction generically does not vanish. We thus predict
\begin{equation}
J - J_{slow} \sim \Gamma_{AB} \label{Ec_slowcorrection}
\end{equation}
for small enough $\Gamma_{AB}$ ($\Gamma_{BA}$) at fixed $\delta$, with
$J_{slow}$ given by (\ref{Ec_slowsolution}).

Similarly, approaching the infinitely fast switching limit a natural anzats is to propose
$P^A (x) \sim P^A_{fast} (x) + \Gamma_{AB}^{-1} W(x)$, with $P^A_{fast}(x)$ given by
(\ref{Ec_fastsolution}). Inserting this ansatz in (\ref{Ec_unasola}) we get the
following result for the fourier-transformed first order correction for $k\neq 0$,
\begin{equation}
-k\sum_{q \neq k} q {\tilde V}_q {\tilde W}_{k-q}-{\chi
|k|^{\alpha}}{ ({\tilde W}_k/\delta)} = -[(\delta^{-1}-1)\chi
|k|^{\alpha}]^2 [{\tilde P^A_{fast}}]_k.
\end{equation}
As before, in the absence of particular symmetries this equation
yields generically solutions with a finite current. This is also
confirmed by direct numerical evaluation of the last equation
for particular potentials, and is also consistent with
the fast switching behaviour of the full numerical
solution presented in the next sections.
We thus predict a non-vanishing first order correction
\begin{equation}
J - J_{fast} \sim \Gamma_{AB}^{-1} \label{Ec_fastcorrection}
\end{equation}
for large enough $\Gamma_{AB}$ ($\Gamma_{BA}$) at fixed $\delta$,
with $J_{fast}$ given by (\ref{Ec_fastsolution}).

It is important to note that the asymptotic behavior of the currents
that we have found (Equations (\ref{Ec_slowcorrection}) and (\ref{Ec_fastcorrection})) are also valid for the case of Gaussian noise ($\alpha=2$). However, they do not coincide with the
corresponding limits found for a randomly flashed triangular
ratchet \cite{astumian_bier} and for a periodically pulsated ratchet
with a frequency $\Omega$ \cite{rmp09}. In this last case, the
corrections analytically found in the slow and fast limits are
$J \sim \Omega^2$ and $J \sim -\Omega^{-2}$ respectively. The
discrepancy is most likely due simply to the difference between a
stochastic and a deterministic switching of the potential.

In the case of \cite{astumian_bier}, however, the discrepancy is more significative
because the system analyzed is also a (dichotomous) stochastically flashed ratchet. Using
an heuristic argument, Astumian and Bier provide an expression for the current as a
function of what we have called $\Gamma_{AB}$ (using $\delta=0.5$), which in the slow and
fast limits behaves as $J \sim \Gamma_{AB}^{3/2}$ and $J \sim \exp(-\Gamma_{AB})$. One of
the possible reasons for this difference could be that that, in the case of a non
analytic potential, the Fourier coefficients do not decay fast enough for our arguments
to be valid. Another possible reason could be that the qualitative argument in
\cite{astumian_bier} is not correct in the limits of fast and slow switching.

\section{Results for a standard ratchet potential}\label{sec_results}

Now we consider a standard ratchet potential \cite{general1}
\begin{equation}\label{Ec_Potencial}
 V(x)=\frac{1}{2\pi}\left[ \sin\left(2\pi x\right)+\frac{1}{4}\sin\left(4\pi
 x\right)\right],
\end{equation}
with period $L=1$ and barriers of amplitude $\Delta V \equiv V_{max}-V_{min} \simeq
0.35$. For this choice of the potential, we analyze the dependence of the current on the
parameters $\chi$ and $\alpha$ characterizing the L\'evy noise, and on the transitions
rates $\Gamma_{AB}$ and $\Gamma_{BA}$. The results of this section correspond, thus, to
the system (\ref{Ec_Langevin}) considering the potential $V(x)$ given in
(\ref{Ec_Potencial}) or, equivalently, to the Fokker-Planck equation
(\ref{Ec_FokkerPlanck}) with $V^A(x)=V(x)$ and $V^B(x)=0$. The results for the current
given throughout this section correspond mostly to the Fokker-Planck formalism. However,
in order to show the complete agreement between the two formalisms we also include
Langevin results for some specific cases.

Concerning the solution of the Fokker-Planck equation, the Fourier
transform of $V(x)$ is simply
\begin{equation}\label{Ec_PotencialFourier}
\tilde{V}_k= \frac{1}{4\pi}
    \left[
              \delta_{k,k+1} + \delta_{k,k-1} +
    \frac{1}{4}\bigl( \delta_{k,k+2} + \delta_{k,k-2}  \bigr)
    \right],
\end{equation}
where $\delta_{i,j}$ is the Kroneker Delta. Thus, the matrix $\mathbf{M}$
given in (\ref{Ec_Matrix}) is penta-diagonal. Despite this
simplification, the system (\ref{Ec_FTFPeqsmat}) can only be
solved numerically, for what we typically consider a number of
Fourier modes ranging from $N=1000$ for $\alpha\sim2$ to $N=2000$
for $\alpha$ close to $1$. This guarantees that our solutions
closely approximate those of the continum limit if the L\'evy noise
intensity satisfy $\chi  > \Gamma_{BA}/|k_{max}|^\alpha$, with
$k_{max}=2\pi N/L$ the maximum wave vector.

\subsection{Symmetrical transition rates}

We first consider the case of equal transitions rates between A and
B states. This means that the potential is `on'and `off' the
same fraction of time, on average. We thus have $\delta = 0.5$ and
we define $\Gamma \equiv\Gamma_{AB}=\Gamma_{BA}$.

\begin{figure}[ht!]
\centering
\includegraphics[width=1.00\textwidth]{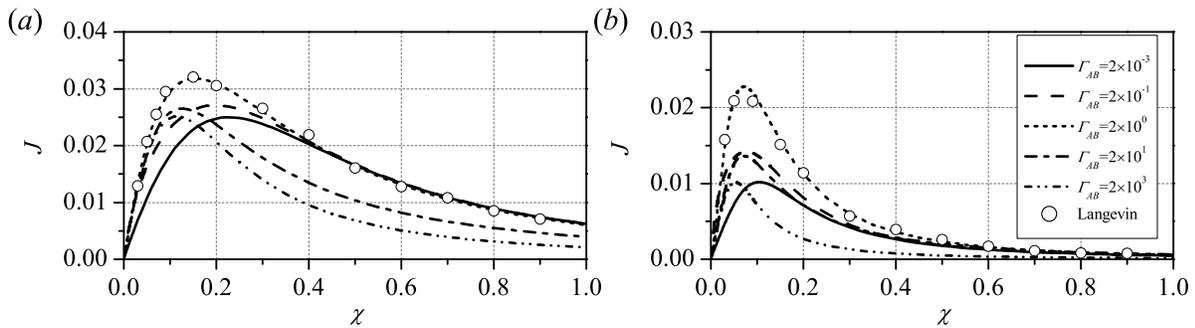}
\caption{Current $J$ as a function of L\'evy noise intensity for
different transition rates ($\Gamma=2\times10^{-3}$,
$\Gamma=2\times10^{-1}$, $\Gamma=2\times10^{0}$,
$\Gamma=2\times10^{1}$, $\Gamma=2\times10^{3}$). $(a)$ stability
index $\alpha=1.4$. $(b)$ stability index $\alpha=1.8$. All the
continuous curves correspond to Fokker-Planck results, while the
open circles were obtained from Langevin simulations for
$\Gamma_{AB}=2\times 10^0$.} \label{Fig_J_vs_Chi}
\end{figure}

In Figure \ref{Fig_J_vs_Chi} we show the current as a function of
the noise power $\chi$ for several values of the transition rate
$\Gamma$ and two different values of $\alpha$. As usual in most
ratchet systems, in all the cases studied we see that the current
attains a maximum at an optimal noise intensity, while it decreases
to zero for low and large intensities. The optimal value of $\chi$
depends only slightly on $\Gamma$. As we can see, it decreases in a
factor of order $1/2$ when changing $\Gamma$ in six orders of
magnitude. We can also see that it decreases with increasing
$\alpha$. In all the cases studied the optimal value of $\chi$
remains in the interval between $\Delta V/10$ and $\Delta V$.

Concerning the weak noise limit $\chi \to 0$, our numerical results
seem to indicate that the current behaves as $J \sim \chi$, in
agreement with the findings for non flashing L\'evy ratchets in
\cite{weaknoise}. However, we have not been able to obtain the exact
analytical limit law within our Fokker-Planck approach.

\begin{figure}[t!]
\centering
  \includegraphics[width=0.95\textwidth]{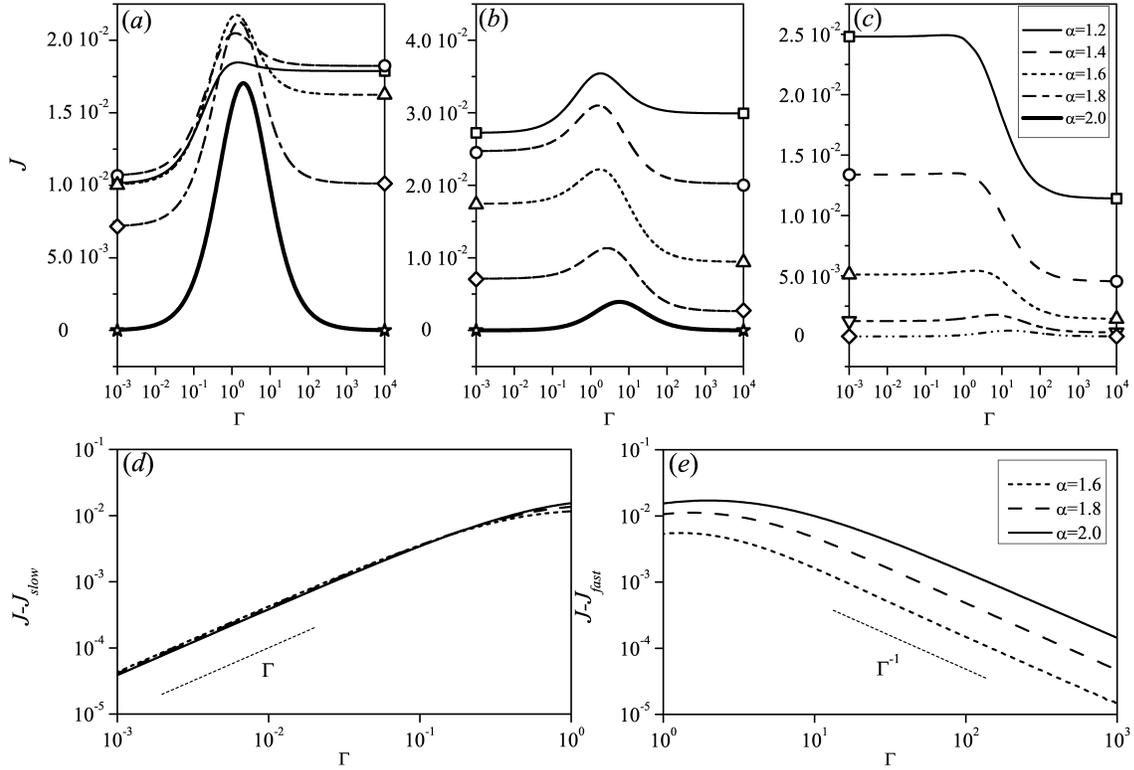}
  \caption{Current as a function of the switching rate for different
values of $\alpha$ considering $\chi=0.05$ $(a)$,$\chi=0.2$ $(b)$ and
$\chi=0.6$ $(c)$. In panels (a), (b) and (c), the symbols at the extreme
values of $\Gamma$ indicate the infinitely slow (Eq.
\ref{Ec_slowsolution}) and infinitely fast (Eq. \ref{Ec_fastsolution})
limit approximations showing complete agreement with the exact
solutions. Panel(d) shows $(J-J_{slow})$ vs. $\Gamma$ in the slow
switching range confirming the validity of the asymptotic formula of
Eq.(\ref{Ec_slowcorrection}). The segment indicates a linear dependence
on $\Gamma$ for reference. Analogously, the results for
$(J-J_{fast})$ in panel (e) shows the validity of the perturbative
analysis at fast switching (Eq.\ref{Ec_fastcorrection}). The segment
indicates a $1/\Gamma$ dependence. Results in panels (d) and (e) are for
$\chi=0.05$.} \label{Fig_J_vs_G}
\end{figure}

Note that, for the two values of $\alpha$ analyzed in Figure \ref{Fig_J_vs_Chi}, we find
that $\Gamma=2$ produces the largest value of $J$ for almost all values of $\chi$, while
larger and smaller transition rates lead almost always to smaller currents. The results
in Figure \ref{Fig_J_vs_G} confirm the existence of an optimum value of $\Gamma$
maximizing $J$, which is almost independent of $\alpha$ and grows with $\chi$.

Figure \ref{Fig_J_vs_G} also shows the validity of the asymptotic formulas found in
section \ref{sec_limitsituations} for the limits of small and large $\Gamma$, and of the
perturbative analysis close to them. Another fact that can be observed in Figure
\ref{Fig_J_vs_G} is that, for a fixed flashing mechanism (fixed $\Gamma$) and a fixed
noise intensity, L\'evy noise leads almost always to larger values of $J$ than Gaussian
noise.

\begin{figure}[ht!]
\centering
    \includegraphics[width=0.95\textwidth]{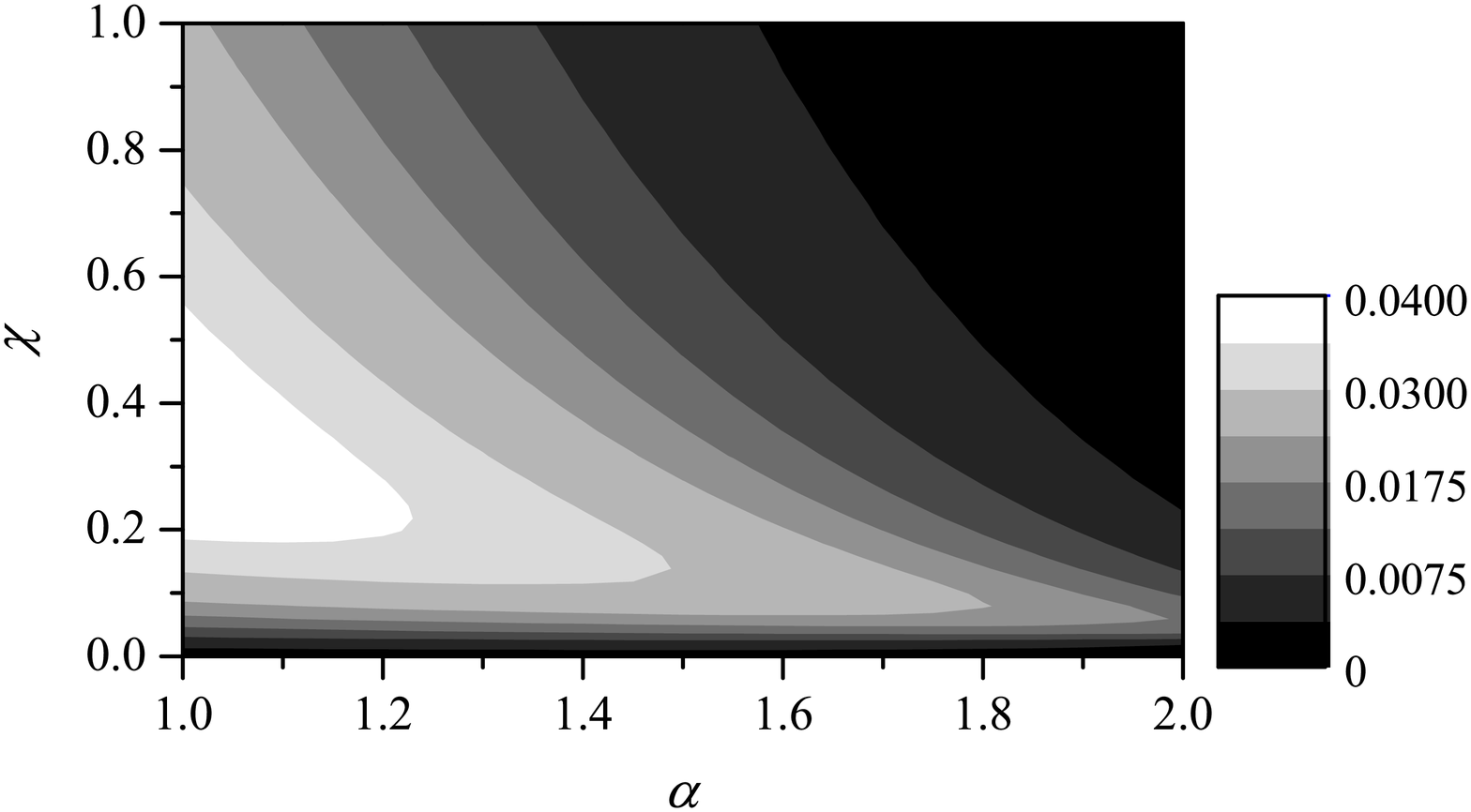}
    \caption{J vs $\alpha$ and $\chi$ for a fixed value of transition rate $\Gamma=2\times 10^{-1}$.}
    \label{Fig_J_vs_AlphaChi}
\end{figure}

Finally, in Figure \ref{Fig_J_vs_AlphaChi} we analyze further the dependence of current
with the noise power and the stability index. In order to identify a relevant noise
region, we plot $J$ as a function of $\alpha$ and $\chi$ for an intermediate (near
optimal) value of the transition rate. For $\alpha$ small, the range of noise intensities
that maximizes the current is centered around $\chi\sim 0.4$. As the stability index
increases, the optimal noise intensity shifts to smaller values. At the same time, the
optimal noise range narrows down as we move to $\alpha = 2$. It is interesting to note
that a given current value can be obtained by combining $\chi$ and $\alpha$ in different
ways. For instance, it is sometimes possible that a small noise power with a large value
of $\alpha$ gives the same current as a large noise power with a small $\alpha$.

\subsection{Non-Symmetrical transition rates}

Now we study the general case of different transitions rates between
`on'and `off' states.

\begin{figure}[t!]
\centering
\includegraphics[width=1.00\textwidth]{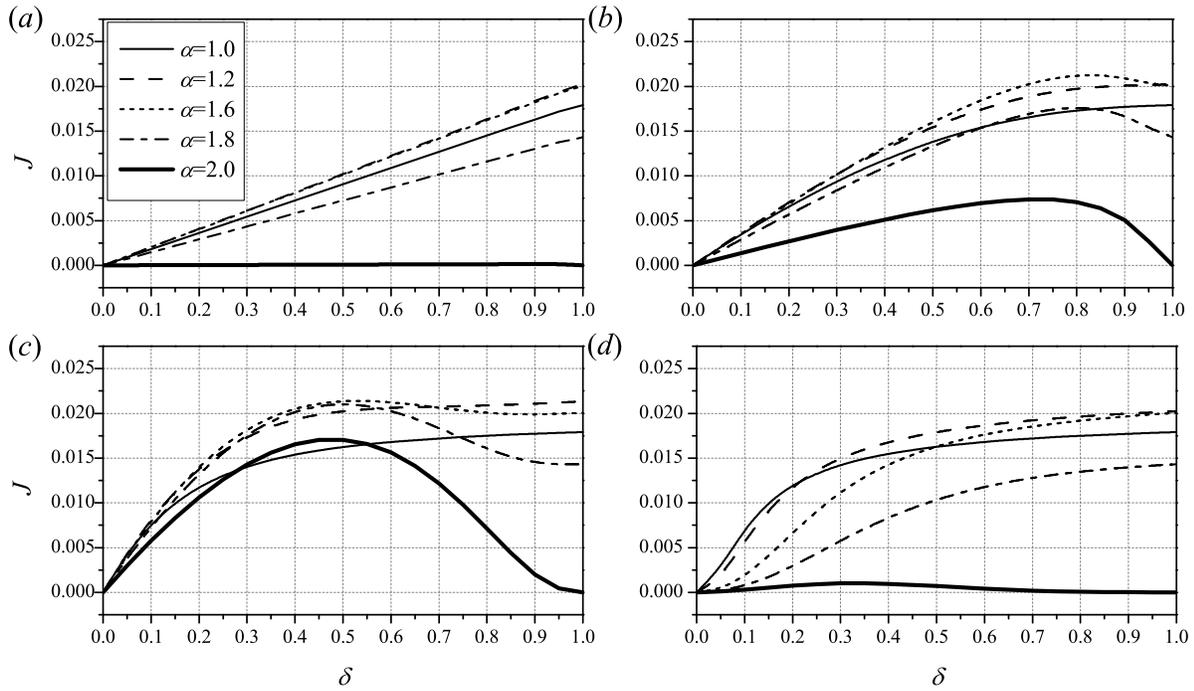}
\caption{Current versus duty ratio for different values of $\alpha$
and fixed $\chi=5\times 10^{-2}$ considering $\Gamma_{AB}=2\times
10^{-3}$ $(a)$, $\Gamma_{AB}=2\times 10^{-1}$ $(b)$, $\Gamma_{AB}=2\times
10^{0}$  $(c)$ and $\Gamma_{AB}=2\times 10^{2}$ $(d)$}.
\label{Fig_J_vs_DR_Chi005}
\end{figure}

In Figure \ref{Fig_J_vs_DR_Chi005} we show the current as a function of $\delta$ for
different values of $\alpha$ and $\Gamma_{AB}$, considering a relatively small noise
intensity $\chi=0.05$. We see that the current vanishes for $\delta \sim 0$. This is
because, at that limit, the potential remains most of the time `off'. In contrast, for
$\delta \rightarrow 1$, the potential is `on'most of the time and we get the static or
non flashing L\'evy ratchet limit. In such situation, the current depends on $\alpha$ and
vanishes only in the case $\alpha \rightarrow 2$, when the system approaches the
equilibrium situation of a non flashing ratchet with Gaussian noise.

The different panels of Figure \ref{Fig_J_vs_DR_Chi005} show us that the transition
probability $\Gamma_{AB}$ plays a significant role in the current behavior (i.e. $\delta$
alone does not determine the dynamics). For intermediate values of $\Gamma_{AB}$ (Figures
\ref{Fig_J_vs_DR_Chi005}$(b)$ and \ref{Fig_J_vs_DR_Chi005}$(c)$) the current has a
maximum at an intermediate duty ratio for almost all values of $\alpha$. In contrast, in
the case of very large or very small ${\Gamma_{AB}}$ (Figures
\ref{Fig_J_vs_DR_Chi005}$(a)$ and \ref{Fig_J_vs_DR_Chi005}$(d)$), except for $\alpha$
equal or very close to $\alpha=2$, the current is a monotonic function of the duty ratio,
and reaches the maximum for $\delta=1$. Note that the linear dependence of $J$ on
$\delta$ observed in figure \ref{Fig_J_vs_DR_Chi005}$(a)$ corresponds to the limit of
slow switching indicated in section \ref{sec_limitsituations}. Namely, $J\simeq \delta \times J_{st}(V)$.

\begin{figure}[t!]
\centering
\includegraphics[width=1.00\textwidth]{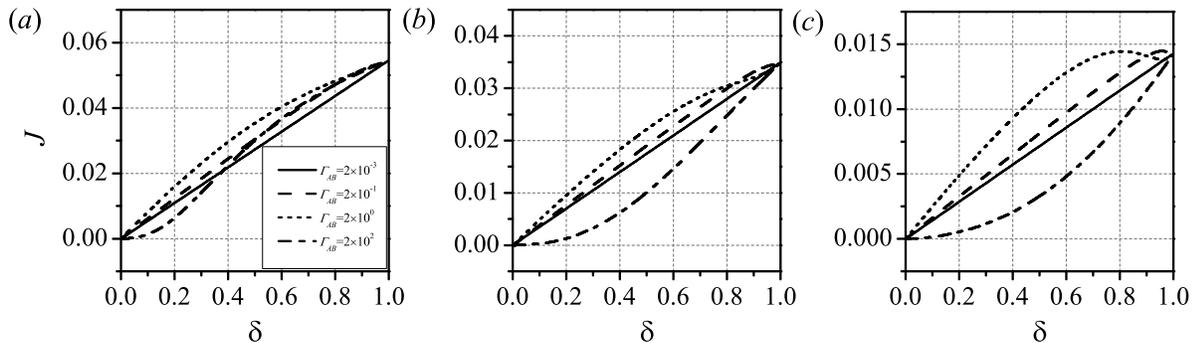}
\caption{Current versus duty ratio for a fixed value of the noise
power $\chi=0.2$ and different values of $\Gamma_{AB}$. Calculations for
$\alpha=1.2$ $(a)$, $\alpha=1.6$ $(b)$, and $\alpha=1.8$ $(c)$.}
\label{Fig_J_vs_DR_Chi020}
\end{figure}

Figure \ref{Fig_J_vs_DR_Chi020} analyzes the dependence of $J$ on
$\delta$ considering a larger noise intensity, and sweeping wide
ranges of $\alpha$ and $\Gamma_{AB}$. We see that in most cases the
maximum current is achieved close to $\delta=1$. The exceptions
occur for values of $\alpha$ close enough to $2$ (for instance that
on Figure \ref{Fig_J_vs_DR_Chi020}$(c)$) and intermediate values of
$\Gamma_{AB}$.

We can summarize the results of our analysis of $J$ as a function of $\delta$ at fixed
$\Gamma_{AB}$ as follows. When considering small enough $\chi$ or large enough $\alpha$,
the inclusion of an appropriate flashing mechanism improves the performance of the static
ratchet. In contrast, for large $\chi$ or small $\alpha$, the largest value of $J$ is
obtained considering a slow flashing mechanism close the static ratchet (i.e.
$\delta=1$). We thus see that the currents depend separately and richly on both,
$\Gamma_{AB}$ and $\delta$, allowing us to both enhance or decrease the static L\'evy
ratchet currents for fixed noise parameters.

\section{Conclusions and final remarks}\label{sec_conclusion}

We have studied the combined action of a L\'evy additive noise and a random dichotomic
flashing in a ratchet system. Our results provide a complete generalization of previous
studies on `non-flashed'  L\'evy ratchets and on standard flashing ratchets with
Gaussian noises.

We have presented a complete analysis of the two-variable fractional
Fokker-Planck equation associated to the system. Our Fourier
treatment allowed us to convert analytically the system of partial
differential equations in an infinite-matrix linear system that can
be easily solved numerically considering an appropriate truncation.
Moreover, we were able to provide analytical asymptotic laws for
slow and fast flashing that behave respectively as $J-J_{slow} \sim
\Gamma$ and $J-J_{fast} \sim 1/\Gamma$, where $\Gamma$ is the
flashing frequency. The solution of the Fokker-Planck equation was
given for an arbitrary periodic potential, and indicated also for
the case in which the system switches randomly between two different
potentials.

Considering a standard ratchet potential we have systematically
investigated the behavior of the current as a function of the
stability index of the L\'evy noise, the noise intensity and the
`on' and `off' rates of the flashing mechanisms. The Fokker-Planck
results for the current were checked by means of Langevin
calculations. We have found that random dichotomic flashing can
produce a rich behaviour of the ratchet current. It allows both to
enhance and diminish appreciably the static L\'evy ratchet current
depending on the magnitude and relative magnitude of the flashing
frequencies, and on the L\'evy noise parameters. A general statement
to remark is that for small enough noise intensity or large enough
stability index, a flashing mechanism can enhance the current of the
static ratchet. Another relevant result indicates that, for a fixed
flashing mechanism, the L\'evy noise gives larger current than the
Gaussian noise in almost any situation.

Our work thus contribute with quite general results and procedures to the understanding
of the transport mechanisms on ratchets. In particular, to the rapid-growing new field of
L\'evy ratchets.

\section*{Acknowledgments}
ABK and SB acknowledge A. Rosso for useful discussions. Support from
CNEA, CONICET under Grant No. PIP11220090100051 and PIP
11220080100076, and ANPCYT under Grant No. PICT2007886 is also
acknowledged.


\section*{References}

\begin{thebibliography}{99}

\bibitem{general1} Reimann P 2002, {\it Phys. Rep.} {\bf 361} 57

\bibitem{rmp09} H\"anggi P and Marchesoni F 2009, {\it Rev. Mod. Phys.} {\bf 81} 387

\bibitem{biology} Chowdhury D, Schadschneider A and Nishinari K 2005, {\it Phys. of Life Reviews} {\bf 2} 318

\bibitem{marchesoni} Savelev S, Marchesoni F, H\"anggi P, and Nori F 2004, {\it Phys. Rev.} E {\bf 70}, 066109; Borromeo M and Marchesoni F 2005, {\it Chaos} {\bf 15} 026110

\bibitem{dybiec} Dybiec B, Gudowska-Nowak E and Sokolov I M 2008,{\it Phys. Rev.} E {\bf 78} 011117

\bibitem{negrete} del-Castillo-Negrete D, Gonchar V Yu and Chechkin A V 2008, {\it Physica} A {\bf 387} 6693

\bibitem{LevyMath} Samorodnitsky G and Taqqu  M S 1994 {\it Stable Non-Gaussian
Random Processes}, (Chapman $\&$ Hall, New York); Applebaum D 2004, {\it L\'evy Processes and Stochastic Calculus}, (Cambridge University Press)

\bibitem{dybiec2} Dybiec B 2008, {\it Phys. Rev.} E {\bf 78} 061120

\bibitem{bouzatPhysA} Bouzat S 2010,{\it Physica} A {\bf 389} 3933

\bibitem{weaknoise} Pavlyukevich I, Dybiec B, Chechkin A V and
Sokolov I M 2010, {\it Eur. Phys. Journ. Special Topics} {\bf 191}
223

\bibitem{negrete2010} Kullberg A, del Castillo Negrete D 2010,
http://arxiv.org/abs/1009.2083.

\bibitem{aclevy} Bao-quan Ai and Ya-feng He 2010, {\it J. Stat. Mech.} P04010

\bibitem{sbudlevy} Bao-quan Ai and Ya-feng He 2010, {\it J. Chem. Phys.} {\bf 132}, 094504

\bibitem{impurity} Vlad M, Spineau F, Benkadda S 2006, {\it Phys. Rev.} Lett. {\bf96} 085001. del-Castillo-Negrete D et. al. 2005 {\it Phys. Rev.} Lett {\bf 94} 065003; Bouzat S et. al. 2006 {\it Phys. Rev.}  Lett. {\bf 97} 205008

\bibitem{phil} Jones P H, Goonasekera M and Renzoni F, {\it Phys. Rev.}  Lett. 2004 {\bf 93} 073904

\bibitem{gommers05a} Gommers R, Douglas P, Bergamini S, Goonasekera M, Jones P H and Renzoni F 2005 {\it Phys. Rev.} Lett. {\bf 94} 143001

\bibitem{gommers05b} Gommers R, Bergamini S and Renzoni F 2005, {\it Phys. Rev.}  Lett. {\bf 95} 073003

\bibitem{quasip}
Gommers R, Denisov S and Renzoni F 2006, {\it Phys. Rev.}  Lett. {\bf 96} 240604;
Gommers R, Brown M and Renzoni F 2007, {\it Phys. Rev.}  A {\bf 75} 053406

\bibitem{gating} Gommers R, Lebedev V, Brown M and Renzoni F 2008, {\it Phys. Rev.}  Lett. {\bf 100} 040603


\bibitem{brown} Brown M and Renzoni F 2008, {\it Phys. Rev.}  A {\bf 77} 033405

\bibitem{kolton} Kolton A B and Renzoni F 2010, {\it Phys. Rev.}  A {\bf 81} 013416

\bibitem{katori} Katori H, Schlipf S and Walther H 1997, {\it Phys. Rev.}  Lett.
{\bf 79} 2221

\bibitem{lutz} Lutz E 2001, {\it Phys. Rev.}  Lett. {\bf 86} 2208; 2004, {\it Phys. Rev.}  Lett. {\bf 93} 190602

\bibitem{Bao} Bao J D 2005, {\it Physica} A {\bf 346} 261

\bibitem{aaa} Horsthemke W and Lefever R 1984, {\it Noise-induced Transitions}, (Springer, Berlin)

\bibitem{bbb} Van den Broeck C, H\"anggi P 1984,{\it Phys. Rev.} A {\bf 30} 2730

\bibitem{astumian_bier} Astumian R D and Bier M 1994, {\it Phys. Rev.}  Lett. {\bf 72} 1766

\bibitem{numeric} Chechkin A V and Gonchar V Yu 2000,{\it Physica} A {\bf 277} 312

\bibitem{podlubny} Podlubny I, {\it Fractional Differential Equations} (Academic Press, London, 1999).

\bibitem{rosso} Zoia A, Rosso A and Kardar M 2007, {\it Phys. Rev.}  E {\bf 76} 021116


\end{thebibliography}
\end{document}